# A three grating electron interferometer

Glen Gronniger, Brett Barwick, and Herman Batelaan

Department of Physics and Astronomy, University of Nebraska—Lincoln, 116 Brace

Laboratory, PO Box 880111, Lincoln, Nebraska 68588-0111, USA

We report the observation of fringes from a three grating electron interferometer. Interference fringes have been observed at low energies ranging from 6 keV to 10 keV. Contrasts of up to 25% are recorded and exceed the maximal contrast of the classical equivalent, Moiré deflectometer. This type of interferometer could serve as a separate beam Mach-Zehnder interferometer for low energy electron interferometry experiments.

Electron interferometry has been applied to many tasks, such as testing the Aharonov-Bohm Effect [1], viewing domain walls in type II superconductors [2], and observing atomic steps in thin films [3]. The breakthrough technology of field emission tips combined with electron bi-prisms lead to the realization of such experiments [4-6]. Proposed, but unrealized experiments for electrons include demonstrating the nondispersive nature of the Aharonov-Bohm effect [7-10] and measuring electron forward scattering amplitude [11]. These and many other experiments such as sensing electric and magnetic fields at surfaces [12, 13], and investigating electron wall decoherence [14-17], are expected too benefit from low energy separate beam interferometry.

The nondispersive nature of the Aharanov-Bohm effect can be shown by pushing the Aharonov-Bohm phase shift beyond the longitudinal coherence length [7-9, 14]. This could be done with a larger solenoid inserted between separated beams at lower electron energies. The cross-section for forward scattering amplitude increases if the energy is lowered into the kV range. To introduce a gas in one interferometer arm a septum has to be inserted between the separate beams [18]. Field sensing due to an electron interacting with surfaces [12, 13] as well as decoherence experiments [14-17] could be enhanced by increasing the interaction times at lower energies.

In atom interferometry a bi-prism interferometer has been developed [19] but most experiments are carried out using grating interferometers [20-22]. Until now only bi-prism interferometers have been available for electron interferometry. If atom interferometry is any indication then there is great promise for a grating interferometer for electrons. It would be exciting to develop electron grating interferometers and investigate their use for the proposed experiments.

In this work we show the first observed fringes from an electron interferometer using nano-fabricated gratings [23]. We observe oscillations in the electron detection rate with a periodicity of about 50 nm and a contrast of maximally 25 %. In principle this observation allows for at least three interpretations. The oscillations could be the result of a quantum mechanical Mach-Zehnder interferometer or Talbot von Laue interferometer, or a classical Moiré deflectometer. All of these devices have useful applications for atoms and molecules [18, 24-26]. The primary concern is to distinguish whether the device is quantum or classical in nature. This distinction allows one to gauge the uses for the device. Secondly, if the device is quantum in nature one can classify it as either near field Talbot von Laue or far field Mach-Zehnder.

For an electron energy of 10 keV, and a grating periodicity of 100 nm the Talbot length, $L_T = d^2/\lambda_{dB}$, is 0.82 mm. This mismatches our grating spacing of 2.54 cm by a factor of about 31, and it is unlikely that we are observing Talbot von Laue fringes. The parameters of our design are chosen so that we reach the Mach-Zehnder domain, i.e., our beam width and beam separation at the second grating are about equal. Larger apparatus length would make the requirements on stray field shielding and alignment more stringent. A slight overlap between the zero and first order diffracted electron beam does not exclude the possibility of weak Moiré or higher order Talbot von Laue fringes. We will show by comparison of our experimental results with both a full quantum mechanical path integral calculation and a classical calculation that we have realized an electron Mach-Zehnder interferometer.

The experimental apparatus is shown in Fig. 1. A slit of 5 µm by 3 mm and a slit of 1.5 ± 0.5 µm by 10 µm separated by 0.24 m, are used to collimate the electron beam produced by thermionic emission using a Kimball Physics EGG-3101 electron gun.

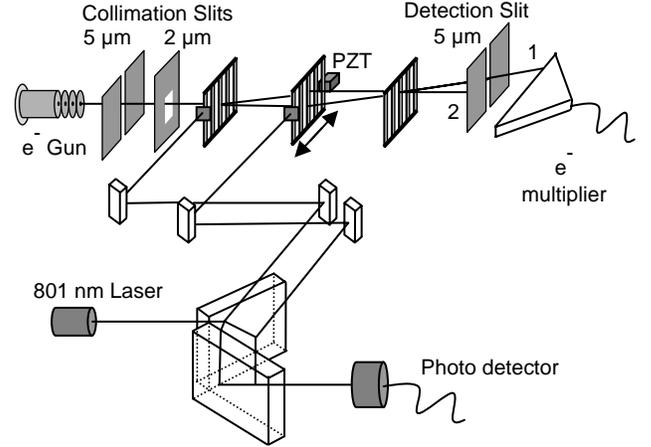

FIG. 1. Sketch showing the experimental setup including the light interferometer used to measure the grating position (Not to scale). Two slits are used to collimate the electron beam before it reaches the three grating (100 nm periodicity) interferometer and an additional slit is used to select the interferometer output port (Output ports 1 and 2 are indicated and 1 is selected in this example).

Our slit configuration gives the best possible beam definition, since we are nearly diffraction limited at the second slit [27]. The distance from the second slit to the first grating of the interferometer is 0.03 m. The 1.2 inch diameter interferometer body is constructed out of titanium. It contains three metal-coated silicon nitride 100 nm periodicity gratings [23] spaced 0.0254 m ± 20 µm between each grating. The middle grating is mounted to a movable slide that is connected to a PZT on one side and fitted with a mirror on the other. The PZT is completely enclosed in titanium. The mirror allows the use of an optical interferometer to measure the movement of the second grating. The fringe contrast dependence on experimental parameters and alignment has been investigated by others for comparable atom interferometers [28]. Rotational alignment was done by observing the diffraction pattern of a HeNe laser from the 1.5 μm period support structure of the gratings. The relative rotational alignment between the gratings is better than 1 mrad. Rotational alignment is not affected by grating motion during an interferometer scan given no loss of contrast in the light interferometer signal. The distance from the third grating in the interferometer to the detection slit is 0.27 m. The 5 μm detection slit is used to select an appropriate output port of the interferometer (Fig. 1). The electrons are detected with an electron channel multiplier. The time independent magnetic fields were shielded to better than 5 mG throughout the vacuum system. The vacuum system is at a pressure of $2 \times 10^{-8}$ Torr, giving us a mean free path much greater than the length of our apparatus. The use of an ion pump and vibrational isolation by an optical table minimizes mechanical noise. With a typical count rate of 200/s and our system parameters we estimate that there is only one electron in the interferometer at any time.

A 801.7 nm New Focus Vortex Laser is used to monitor the position of the second grating. Two parallel beams from a partially monolithic Michelson interferometer are reflected from two separate mirrors. One mirror is connected to the moveable second grating and the other is connected to the body of the interferometer. The interference signal from the light interferometer is collected along with the PZT ramp signal from the saw tooth wave of a function generator. This allows us to simultaneously take monitoring data along with the electron interferometer signal. The drift and vibrational motion of the second grating relative to the interferometer body does not exceed 10 nm for all data runs. Drifts of 10 nm are estimated to reduce the observed contrast by approximately 2%.

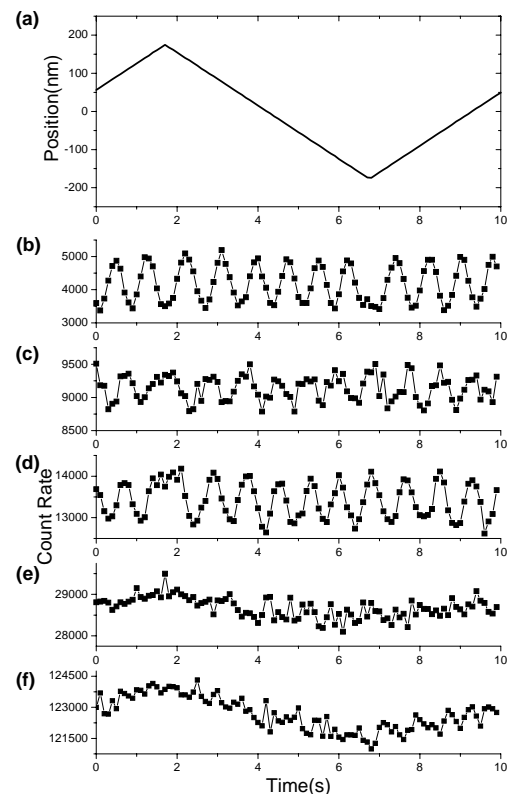

FIG. 2. Electron interference data at output port 1. Fig. 2a is the relative position of the second grating in nanometers. The experimental data is given for (b) 10 keV, (c) 8 keV, (d) 6 keV, (e) 4 keV, and (f) 2 keV. The 10, 8, and 6 keV data show 50 nm periodicity fringes, while the 4 and 2 keV do not show fringes.

Fringes have been observed for energies ranging from 6 keV to 10 keV

but not at 2 keV and 4 keV (Fig. 2). This data was taken at output port number 1 (Fig. 1). The time axis in Figure 2 represents one full grating scan sweep. The electron count rate data is the sum of multiple sweeps. The lack of fringes at the lower energies is not unexpected. Any stray or PZT fields become prominent at lower energies. Stray fields and patch fields [29, 30] can effect the longitudinal and transverse coherence of the electrons. At even lower energies (500 eV to 50 eV) [27] the grating structure can cause dephasing. These problems can be overcome. Longitudinal phase shifts between interferometer arms can be compensated with the introduction of a Wien filter [31]. Patch fields can be reduced by increasing the interferometer bore. Stray magnetic and PZT fields can be suppressed with better shielding. A different choice of metallic coating on the gratings can reduce this grating dephasing [27]. With such measures we expect that the interferometer can be operated to below 1 keV.

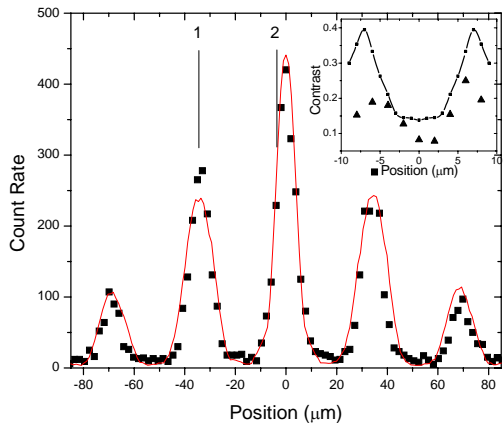

FIG. 3. Electron diffraction through the interferometer at 10 keV. Solid line is the full path integral quantum mechanical calculation. Interferometer output ports are numbered 1 and 2. Contrast as a function of detector position for the zero order is shown in the inset (triangular data points). The solid line in the inset corresponds to a spline fit through the theoretical data points. The dip in contrast is a feature of a Mach-Zehnder interferometer.

The diffraction through the interferometer shown in Fig. 3 is used to identify the output ports of the interferometer. This allows us to place our detection slit at one of the output ports. First order diffraction peaks from individual gratings are not resolved but zero and first orders are resolved. The diffraction pattern agrees well with a path integral calculation without any interactions between the electron and the grating bars [32]. Maxima in the contrast as a function of detector position are found around the 0 order (Fig. 3 inset). The dip in the fringe contrast at the 0 order in the experimental data is in agreement with calculation and is a characteristic of the Mach-Zehnder interferometer.

We observe a periodicity of 50 nm while fringes would have 100 nm periodicity at integer multiples of the Talbot length. This excludes the possibility that we are observing such fringes. Classical Moiré fringes do have the same period as the fringes that we observe. For our experimental parameters we performed a Moiré deflectometer simulation which yields a maximum contrast of 5%. Experimentally we observe maximum contrasts of 25% which excludes the Moiré deflectometer explanation, Fig 4. The quantum mechanical path integral calculation gives a contrast of about 15 - 40 % depending on the detector position, Fig 4. This is always somewhat larger than the experimental contrast at the same interferometer output port. This is reasonable given some reduction of contrast due to slight misalignments.

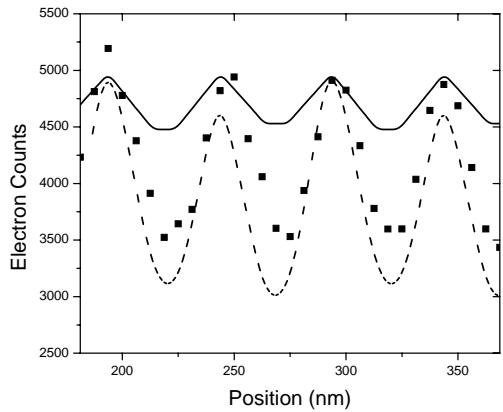

FIG. 4. Experimental data comparison to theory. The result of the classical straight line path calculation is represented by the solid line. The result of the full path integral calculation is represented by the dashed line. Experimental data is represented by square dots. Contrast of our device exceeds the classical contrast by about 3 times showing the quantum mechanical nature of our data.

Our results show that electron grating interferometry is possible. The ability to go to lower energies has its difficulties but it is likely that all of the mentioned low energy problems can be overcome. The use of this device to probe fundamental physics such as testing the nondispersive nature of the Aharonov-Bohm effect as well as other proposed experiments is exciting. Based on the success of atomic, molecular, and neutron interferometers constructed from gratings [7, 20, 22, 25, 26, 33, 34], we feel that it is important to investigate electron grating interferometers further.

## Acknowledgements

We thank Adam Caprez for help with the theoretical calculations, as well as David Swanson and The University of Nebraska-Lincoln Research Computing Facility. We also thank Stephanie Gilbert for her help with the monitoring light interferometer. This material is based upon work supported by the National Science Foundation under Grant No. 0112578. This material is also based upon work supported by the Department of the Army under Grant No. DAAD1902-1-0280, and the content of the information does not necessarily reflect the position or the policy of the federal government, and no official endorsement should be inferred.

Correspondence and requests for materials should be addressed to H. B. (e-mail: hbatelaan2@unl.edu).